\documentclass[rnote]{aa}
\usepackage{amsmath,graphicx,amssymb}
\usepackage[varg]{txfonts}

\newcommand{\Ondrejov}{Ond\v{r}ejov}
\newcommand{\Halpha}{H$\alpha$}
\newcommand{\Pbeta}{P$\beta$}
\newcommand{\Brgamma}{Br$\gamma$}
\newcommand{\kms}{\,\mbox{$\mathrm{km}~\mathrm{s}^{-1}$}}

\newcommand{\jednadel}{1\,\,Del}
\newcommand{\jednadela}{{\jednadel}\,A}
\newcommand{\jednadelb}{{\jednadel}\,B}
\newcommand{\BDjednadel}{BD\,+10$^{\circ}$4303}

\newcommand{\SINFONI}{SINFONI}
\newcommand{\Qfitsview}{{\tt QFitsView}}

\begin{document}

\title{Spectroscopy of close visual binary components of the stable
shell star {\jednadel}%
\thanks{Based on observations collected with the Perek \mbox{2-m}
Telescope at the {\Ondrejov} Observatory, Czech Republic and the
{\SINFONI} spectrograph at UT4-Yepun in ESO during the observing
programme \mbox{093.D-0172}.}}

\author{
Ji\v{r}\'{\i} Kub\'at\inst{\ref{ondrejov}}
\and
Brankica Kub\'atov\'a\inst{\ref{ondrejov},\ref{sanu}}
\and
Barbora Dole\v{z}alov\'a\inst{\ref{ondrejov},\ref{brno}}
\and
Lubomir Iliev\inst{\ref{sofia}}
\and
Miroslav \v{S}lechta\inst{\ref{ondrejov}}
}

\authorrunning{J. Kub\'at et al.}

\institute{
	Astronomick\'y \'ustav, Akademie v\v{e}d \v{C}esk\'e republiky,
	CZ-251 65 \Ondrejov, Czech Republic
	\label{ondrejov}
	\and
	Matemati\v{c}ki Institut SANU, Kneza Mihaila 36, 11001 Beograd,
	Serbia\label{sanu}
	\and
	\'Ustav teoretick\'e fyziky a astrofyziky P\v{r}F MU,
	Kotl\'a\v{r}sk\'a 2, CZ-611 37 Brno, Czech Republic
	\label{brno}
	\and
	Institute of Astronomy, Bulgarian Academy of Sciences, 72
	Tsarigradsko Shossee Blvd., BG-1784 Sofia, Bulgaria
	\label{sofia}
}

\date{Received 26 April 2015 / Accepted 3 December 2015}

\abstract{Stable shell stars are ideal objects for studying basic
physical principles of the formation of disks in Be stars.
If these stars have a close unresolved visual companion, its
contribution toward the modelling of the disk cannot be overlooked, as
is sometimes done.
The study aims to spectroscopically resolve close visual binary Be
(shell) star {\jednadel}, which up to now was only resolved  by speckle
or micrometric measurements.
The integral field spectroscopy obtained by the {\SINFONI} spectrograph
at the VLT telescope in the European Southern Observatory (ESO) in the
infrared region was used; we supplemented these observations with visual
spectroscopy with the Perek Telescope at the {\Ondrejov} Observatory.
Spectra of {\jednadel} were successfully resolved, and,
for the first time, spectra of {\jednadelb} were obtained.
We found that {\jednadela} is a Be/shell star, while {\jednadelb} is
not an emission-line object.}

\keywords {stars: emission-line, Be
	-- binaries: visual
	-- infrared: stars
	-- stars: individual: {\jednadel}}

\maketitle

%%%%%%%%%%%%%%%%%%%%%%%%%%%%%%%%%%%%%%%%%%%%%%%%%%%%%%%%%%%%%%%%%%%%%%%%
\section{Introduction}

Be~stars form an important subgroup of stars of spectral class~B.
They are defined as non-supergiant B-type stars, which exhibit, or at
some time in their observational history have exhibited, emission
superimposed over absorption lines in their spectra
\citep{Collins:1987}.
This emission is most commonly found in the {\Halpha} line of the Balmer
series of hydrogen.
Shell stars form a subclass of Be stars \citep[see e.g.][]
{Hanuschik:1996}.
They are characterized by sharp absorption cores of rotationally
broadened emission lines.
These stars are often variable, for example the star Pleione
\citep[see][]{be27}.
Stable shell stars form their special subclass \cite[see][]
{Gulliver:1981}.
Their stability makes them especially suitable for testing basic
physical processes that may lead to the existence of a disk, since the time
dependent phenomena do not complicate the analysis.
An understanding of stable shell stars may shed light on Be and shell stars
with much more complex and time dependent behaviour.
For recent reviews about Be stars and their models, see
\cite{Porter:Rivinius:2003} and \cite{Rivinius:etal:2013}.

The importance of stable emission-line stars for developing basic
physical models of emission formation was already noticed by
\cite{Marlborough:Cowley:1974}, who used the stable shell star
{\jednadel} as a typical stable emission-line star for modelling.
More sophisticated models were later calculated by
\cite{Millar:Marlborough:1999} and \cite{Jones:etal:2004}.
These authors were able to reproduce the profiles of the observed
{\Halpha} line profile of {\jednadel} with their model of the
circumstellar envelope.
They assumed that {\jednadel} is a single star.

However, {\jednadel} was discovered as a visual binary with
components separated by less than one arcsecond \citep{Burnham:1873}.
Also, many other Be stars are in close visual binaries
\cite[see][]{Oudmaijer:Parr:2010}.
Spectroscopic observations of these close visual binaries is not an easy
task, since many times the observing conditions are not perfect and
light from both stars overlaps.
The small separation of the components of the close pair very often
causes these stars to appear at the spectrograph slit as a single
object.
As a consequence, it is difficult or impossible to resolve individual
spectra.
Observed spectra contain mixed light from both stars.
If the magnitude difference between components is not large, both stars
significantly contribute to observed spectra and this fact has to be
taken into account in analysis using single star disk models.
In this paper, we aim to extract spectra of individual components of
{\jednadel}.

%%%%%%%%%%%%%%%%%%%%%%%%%%%%%%%%%%%%%%%%%%%%%%%%%%%%%%%%%%%%%%%%%%%%%%%%
\section{Summary of basic facts about the star \jednadel}

The star \object{\jednadel} (HR\,\,7836, HD\,\,195325, \BDjednadel,
HIP\,\,101160) belongs to the category of above mentioned stable
emission-line shell stars in close visual binaries.

The shell spectrum of {\jednadel} was first detected on 9 August 1919 at
the Dominion Astrophysical Observatory \citep[see][]{Bidelman:1988}.
Numerous emission lines were reported by \citet[][observations from
1929]{Harper:1937} and \cite{Bidelman:1949}.
\citeauthor{Bidelman:1949} also noted that the primary component of the
visual binary is a shell star.
The star {\jednadel}
is also listed by \citet[MWC\,1019]{Merrill:Burwell:1949} as an
emission-line star.
This star is classified as  stable, since no change of its spectrum
was recorded between 1953 and 1980 \citep[see][] {Gulliver:1981}.
Variability found by \cite{Abt:2008} and \cite{1del} is very small.
Several authors determined the spectral type of {\jednadel}, which
varies between B8 and A1, the most recent determination is A1:III\,shell
\citep{Abt:Morrell:1995}.

{\jednadel} is a triple visual system (ADS\,13920, CCDM J20303+1054).
\cite{Burnham:1873:areg} reported this star as a hardly observable visual
binary.
His measurements were published in \cite{Burnham:1873} as  star No.63
(now denoted as BU\,\,63).
Later \citet{Burnham:1874} found a faint companion to this binary (star
No.297, BU\,\,297).
This and later position measurements of components of {\jednadel} are
listed in Table~\ref{polohy1del} (online).
To summarize, this triple visual system consists of a close visual
binary, {\jednadela} ($V=6.1$) plus {\jednadelb} ($V=8.1$), separated by
$0.9\arcsec$, and a more distant fainter star {\BDjednadel}\,C
($V=14.1$, $\rho=16.8\arcsec$).
These values of magnitudes, separation, and angle are taken from
\cite{Dommanget:Nys:2000}.

%%%%%%%%%%%%%%%%%%%%%%%%%%%%%%%%%%%%%%%%%%%%%%%%%%%%%%%%%%%%%%%%%%%%%%%%
\section{Observations and data reduction}

%=======================================================================
\subsection{{\Ondrejov} visual spectra}

Visual spectra were observed with the Perek \mbox{2-m} telescope at the
{\Ondrejov} Observatory (Czech Republic) with the slit spectrograph in
coud\'e focus \citep[for description, see][]{Slechta:Skoda:2002}.
We obtained medium-resolution spectra, ($R\sim 13\,000$) with the 700-mm
spectrograph focus, in which the SITe CCD 2030$\times$800 chip of
15$\mu$m pixels was used.
Our spectra of {\jednadel} were taken on 16 September 2007.
The data were reduced (bias subtracted, flat-fielded, and wavelength
calibrated) using the Image Reduction and Analysis Facility
(IRAF\footnote{IRAF is distributed by the National Optical Astronomy
    Ob\-ser\-va\-to\-ries, which are operated by the Association of
    Universities for Research in Astronomy, Inc., under cooperative
    agreement with the National Science Foundation.}).
The spectra were normalized to continuum with a straight line to fit the
far wings of the {\Halpha} line.

%=======================================================================
\subsection{Infrared {\SINFONI} spectra}

The observations were obtained in service mode at the VLT-UT4 (Yepun)
telescope at ESO under Programme-ID 093.D-0172(A) during two nights (6
and 10 June 2014).
Spectroscopy of {\jednadel} was carried out with the SINFONI, the
AO-assisted integral field spectrograph
\citep{Eisenhauer:etal:2003,Bonnet:etal:2004} in all available
near-infrared bands (J, H, and K).

The integral field unit of the SINFONI spectrograph divides the field of
view to surface elements (usually referred to as ``spaxels'').
The spatial scale was set to $0.1\arcsec$ (``spaxel'' scale of
$50\,\mathrm{mas} \times 100\,\mathrm{mas}$) per pixel with the field of
view of $3\arcsec \times 3\arcsec$.
Calibrations (dark frames, flatfield frames, and the telluric standard
star) were provided by the ESO baseline calibration.
The image was rotated using position data from Table \ref{polohy1del} to
make the line connecting the close components parallel with the boundary
of the field of view.
During observations the seeing was between $0.6\arcsec$ and
$2.7\arcsec$.

We reduced the raw data  via the {\SINFONI} pipeline version 2.4.0.
Standard reduction steps were taken to the final product of 3D datacubes
with one spectral and two spatials
dimensions.
We used the programme {\Qfitsview}%
\footnote{\url{http://www.mpe.mpg.de/~ott/QFitsView/}} for visualization
of 3D spectra in image cubes and spectrum extraction.
The extracted spectra were normalized to continuum with a straight line.

%%%%%%%%%%%%%%%%%%%%%%%%%%%%%%%%%%%%%%%%%%%%%%%%%%%%%%%%%%%%%%%%%%%%%%%%
\section{Identification of binary components}

%=======================================================================
\subsection{Test of close binary visual spectra}
\label{testjednadel}

\begin{figure}
\includegraphics[width=\hsize]{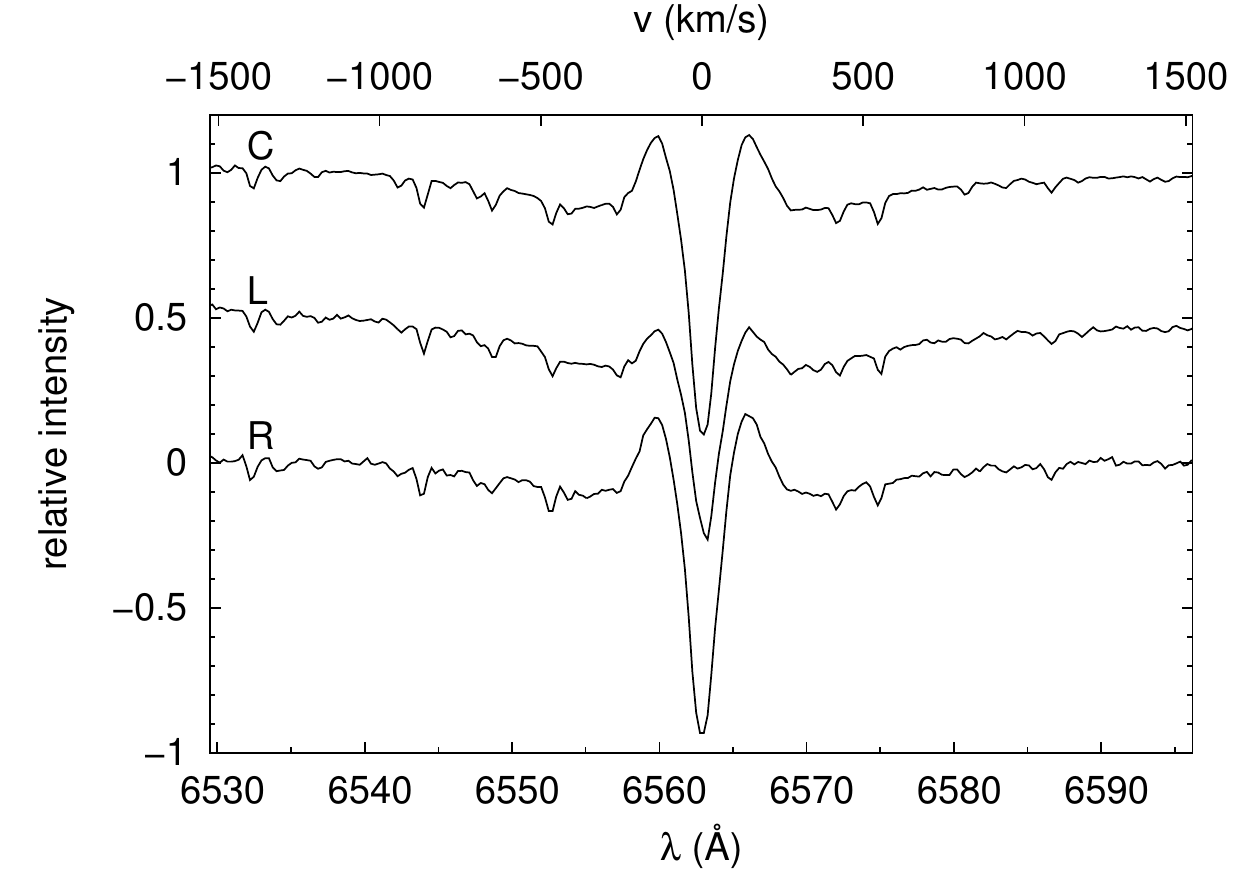}
\includegraphics[width=\hsize]{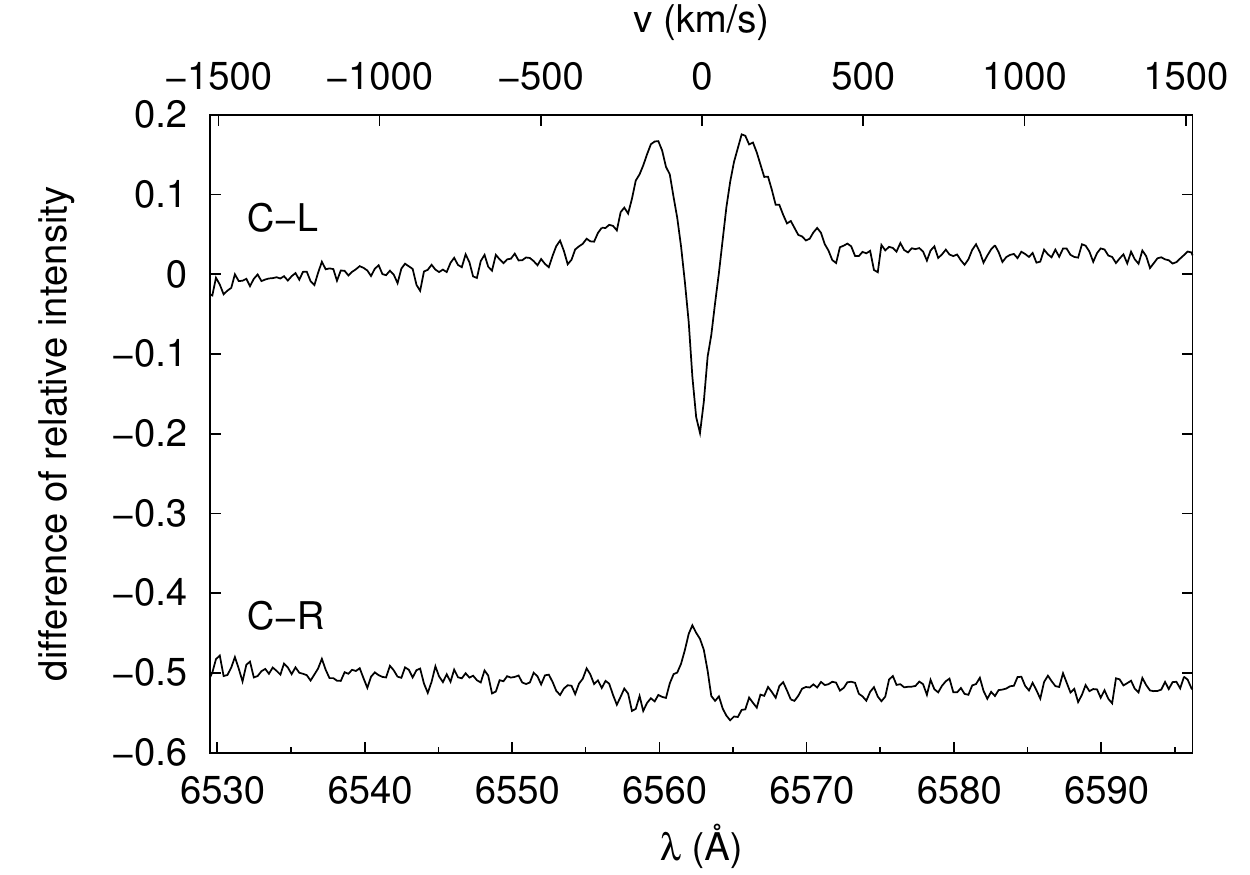}
\caption{{\bf\em Upper panel:}
Normalized spectra of the H$\alpha$ line of {\jednadel} obtained at
different places of a stellar image.
Spectra were observed one after another during one night.
They were obtained by positioning the slit to the left (L), central (C),
and right (R) part of the stellar image.
In this plot, spectra of (L) and (R) are shifted down by 0.5 and 1.0,
respectively.
{\bf\em Lower panel:}
Differences between the central (C) spectrum and the left (L) and right
(R) spectra.
The difference (C$-$R) is shifted down by 0.5.
Velocities in both panels are expressed with respect to the
centre of the {\Halpha} line.}
\label{1delhaond}
\end{figure}

Obtaining spectra of individual components of a visual binary when
images of the stars overlap is a challenging task.
Consequently, the observed spectrum contains spectra of both components.
This is also the case of observations at the {\Ondrejov} Observatory,
where even the best seeing conditions are of the order of arcseconds,
thus, larger than the apparent separation of the visual components of
both binaries.
They appear only as one object on a spectrograph slit.

Nevertheless, we carried out the following experiment with the binary
{\jednadel}.
Since the Perek 2-m telescope is not equipped with the derotator, the
right moment when the connecting line between the A and B components was
roughly perpendicular to the slit had to be chosen.
Three spectra in the region of the {\Halpha} line were secured, one
after another.
The $0.2\arcsec$ wide slit was placed at three different parts of the
stellar image at the slit camera, at the left hand side (L), right hand
side (R), and at the centre (C) of the stellar image.
All three exposures were 900\,seconds long.
The raw signal was highest for C as expected, but the raw signal for R
was about seven times fainter than for L.
This means that the fainter component (\jednadelb) was located at or
near L at the time of observation.

Normalized spectra for C, R, and L are shown at the upper panel of
Fig.~\ref{1delhaond}.
The emission height with respect to the local continuum is
smaller for L than for R.
This corresponds to the case when additional continuum (without an
emission) is added to the radiation of the star, and the emission strength
weakens.
All this indicates that the {\Halpha} emission should come from the
brighter component, namely {\jednadela}, rather than from the fainter
component, {\jednadelb}.
In the difference spectrum C$-$L (see the lower panel of
Fig.~\ref{1delhaond}), the emission in the {\Halpha} line is stronger
than in the C spectrum.
This  shows that the spectrum of {\jednadela} is an emission spectrum
and that the spectrum of {\jednadelb} is either an absorption spectrum
or it has a weak emission, which appears in the difference spectrum
(C$-$R).
However, these observations involve strong uncertainties (mainly because
of uncertain positions of components A and B during observations), thus,
it is not possible to reach a firm conclusion.
To obtain more information about the system, we decided to observe
the binaries with an integral field spectroscopy device, which is
available in the infrared spectral region.

%=======================================================================
\subsection{Infrared integral field spectroscopy}

\begin{figure}
\centering
\includegraphics[width=\hsize]{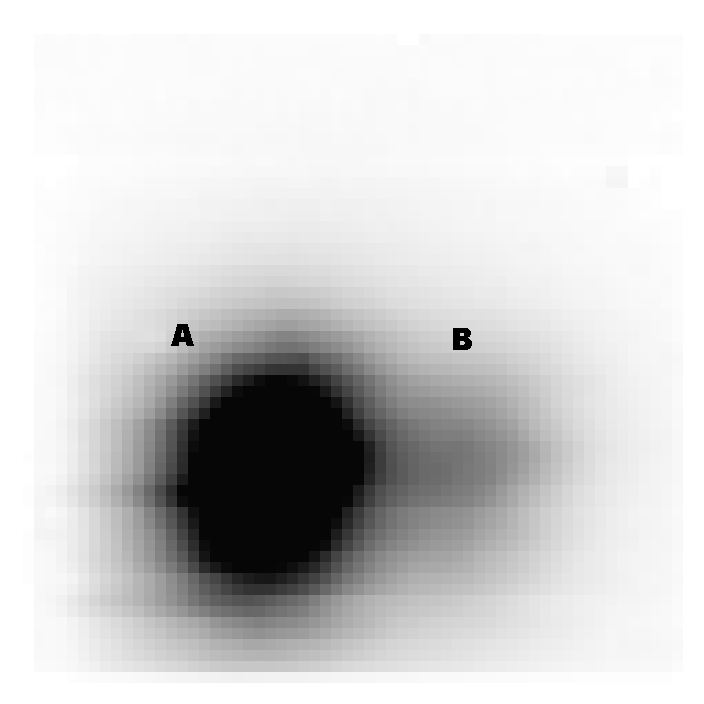}
\caption{Image of the {\jednadel} binary obtained by {\Qfitsview} in the
J~band.
Both stars are seen clearly.
The A component is in the left part of the figure (denoted by the letter
A); the B component is in the right part of the figure (denoted by B).}
\label{1Delobrazek}
\end{figure}

\begin{figure}
\includegraphics[width=\hsize]{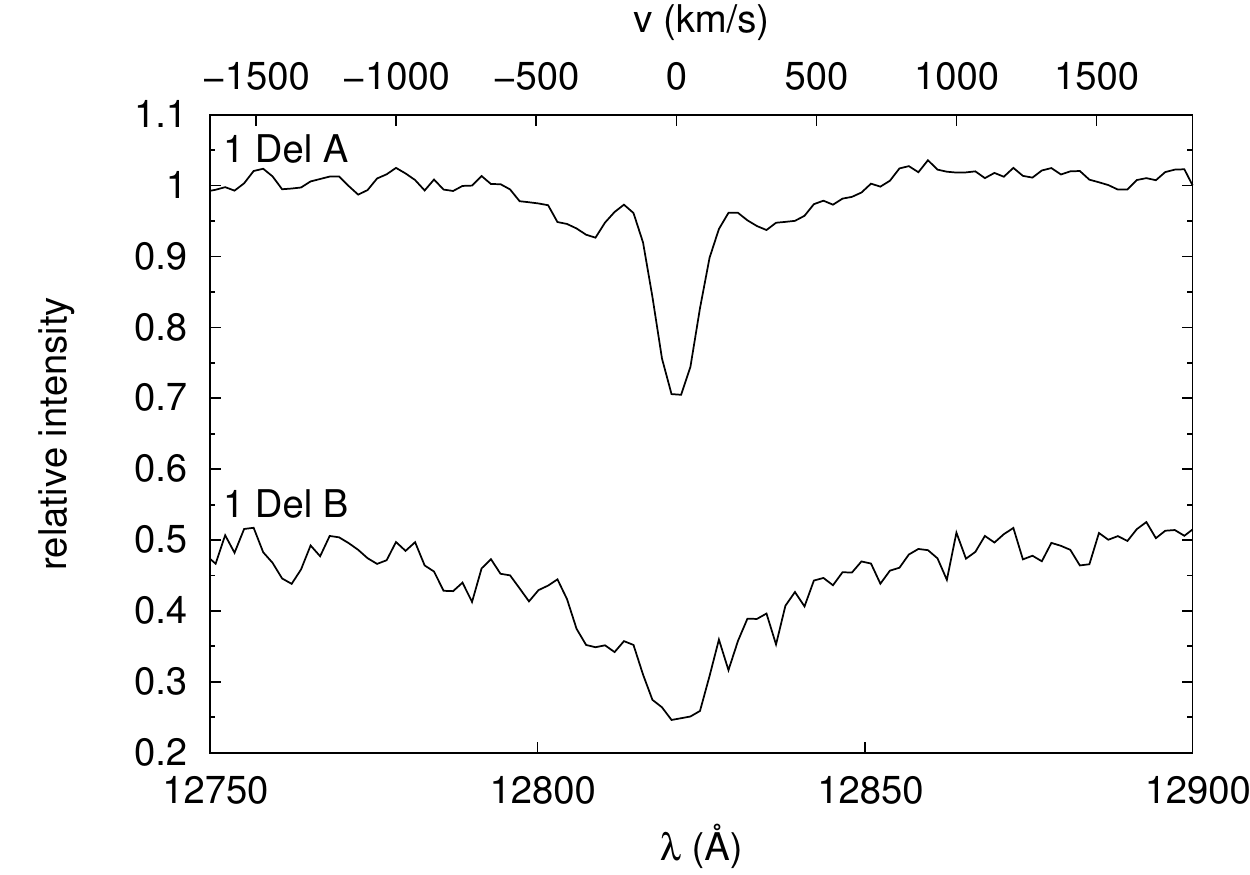}
\caption{Normalized spectra of A and B components of {\jednadel} in the
region around  {\Pbeta} line.
The spectrum of {\jednadelb} is shifted down by 0.5.
Velocities are expressed with respect to the centre of the
{\Pbeta} line.}
\label{pbrgam}
\end{figure}

\begin{figure*}
\includegraphics[width=\hsize]{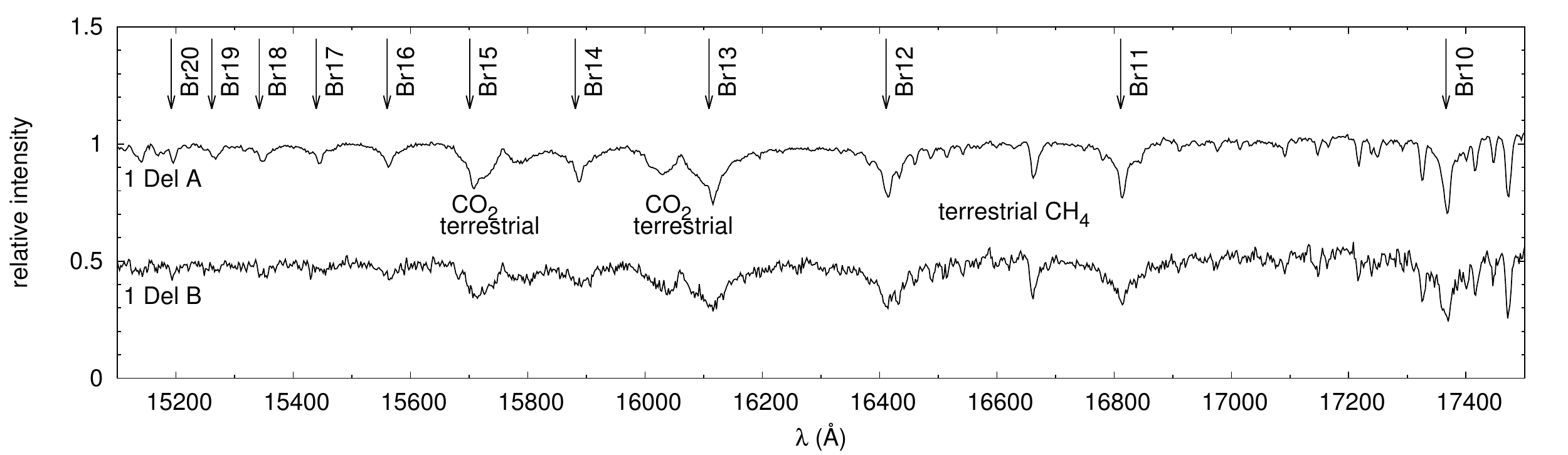}
\caption{Normalized spectra of A and B components of {\jednadel} in the
region of higher Brackett lines.
The spectrum of {\jednadelb} is shifted down by 0.5.}
\label{brlin}
\end{figure*}

The image of the binary {\jednadel}, obtained by the programme {\Qfitsview}
from the observations in the J~band, is given in
Fig.~\ref{1Delobrazek}.
The secondary (\jednadelb) is clearly visible in the right part of the
image.
Stellar spectra were extracted using the programme {\Qfitsview}, which
enables extraction of spectra for individual spaxels or integrated over
a circle of spaxels.
The light was integrated from the regions occupied by A and B components
with a minimized contribution of the other component.
This way we obtained the spectrum of {\jednadela} and the very first
spectrum of {\jednadelb}.

The spectra in the region of the {\Pbeta} line are plotted in
Fig.\,\ref{pbrgam}, where the difference between spectra of {\jednadela}
and {\jednadelb} is clearly visible.
While there is no emission in the {\Pbeta} line of {\jednadelb}, the
shell features in the wings of the {\Pbeta} line are clearly seen in
the spectrum of {\jednadela}.
Consequently, the A component is a shell star,
as indicated by rough estimates from the {\Halpha} line profiles.
The higher hydrogen Brackett series lines are in the upper panel of
Fig.\,\ref{brlin}.
In this region, the absorption in the lines of {\jednadela} is sharper,
confirming this component as a shell star.
The highest members of the Brackett series (above Br17) are hardly seen
in the spectrum of {\jednadelb}, while they are clearly visible for
{\jednadela}.
The strong absorption line near $16700\,\AA$ is caused by a vibrational
transition of terrestrial methane \citep[see e.g.][]{Brown:etal:2003},
which also causes other surrounding small absorption lines.
The strong terrestrial, vibrational absorption of $\text{CO}_2$ \citep[see
e.g.][] {Rothman:etal:2005, Tashkun:etal:2015} blends with Br13 and
Br15 lines.
The {\Brgamma} line (not shown here) also has emission with a shell
absorption core for the A component,
however, the spectra in this spectral region are very noisy.

To summarize, the observations show that only the A component of
{\jednadel} is an emission-line star.
The B component seems to be of a similar spectral type as the A
component (without an emission) with significantly broadened absorption
lines.
By  comparing H-band spectra of Be stars presented by
\cite{Steele:Clark:2001}, we may classify {\jednadelb} as late B.
Unfortunately, the H-band region is not suitable for more exact spectral
type determination or  for rotational velocity determination, since
this region lacks uncontaminated photospheric features
\citep[cf.][]{Steele:Clark:2001}.
Nevertheless, we tried to estimate the rotational velocity of
{\jednadelb} from the Br11 line.
We applied a technique used, for example by \cite{Slettebak:1967}, for
A-type stars in the visual spectral region, and obtained 370\kms, which
is a larger value than for {\jednadela}.
The most recent value by \citealt{Royer:etal:2002} is 217\kms.
Since the apparent brightness difference between A and B components is
only about 2 magnitudes, then if the components
are not resolved, radiation from the B component has to be taken
into account in fitting the spectrum of {\jednadel}.

%%%%%%%%%%%%%%%%%%%%%%%%%%%%%%%%%%%%%%%%%%%%%%%%%%%%%%%%%%%%%%%%%%%%%%%%
\section{Conclusions}

Resolved spectra of both components of the stable shell star {\jednadel}
were obtained for the first time with infrared integral field spectra.
We showed that {\jednadela} is a shell star, while {\jednadelb} is
not an emission-line star.

\begin{acknowledgements}
This research was supported by GA\,\v{C}R 13-10589S.
This research has made use of the Washington Double Star Catalog
maintained at the U.S. Naval Observatory.
The authors would like to thank the anonymous referee for his/her
comments on the manuscript, which led to significant improvement of the
paper.
\end{acknowledgements}

\newcommand{\an}{Astron. Nachr.}

\bibliographystyle{aa} % style aa.bst
\bibliography{shbin,kubat} % your references Yourfile.bib

\Online

\begin{sidewaystable*}[h]
\caption{Measurements of angular separation of components of the triple
system {\jednadel}.}
\label{polohy1del}
\begin{center}
\begin{tabular}{|l|c|c|c|c|c|c|c|c|c|c|l|}
\hline
date & \multicolumn{2}{|c|}{angle} & \multicolumn{2}{|c|}{separation} & $n$ & \multicolumn{5}{|c|}{magnitudes} &
	reference
\\
& \multicolumn{2}{|c|}{[$^\circ$]} &
\multicolumn{2}{|c|}{[$^{\prime\prime}$]} & & \multicolumn{5}{|c|}{} &
\\
\cline{2-5}
\cline{7-11}
& AB & AC & AB & AC & & A & B & A$-$B & C & A$-$C & \\
\hline
4.10.1872 & 340 & & 0.7 & & & 6 & 11 & & & & B73 \\
\hline
1874.54 & 343 & & 0.85 & & & & & & & & B74\tablefootmark{D} \\
        & & 360 & & 15. & & & & & & & B74 \\
\hline
1948.74 & 348.3 & & 0.85 & & & 6.0 & 8.0 & & & & P51 \\
1948.76 & 351.9 & & 0.75 & & & & & & & & \\
1948.77 & 349.9 & & 0.83 & & & & & & & & \\
$\mathit{1948.76}$ & $\mathit{350.0}$ & & $\mathit{0.81}$ & & $\mathit{3}$ &
	& & & & & {\em m} \\
\hline
1957.475 & 345.9 & & 0.97 & & & 6.0 & 8.0 & & & & vdB58 \\
1957.490 & 349.4 & & 0.92 & & & 6.0 & 8.2 & & & & \\
1957.498 & 345.7 & & 0.86 & & & 6.0 & 8.2 & & & & \\
1957.512 & 345.7 & & 0.89 & & & 6.0 & 8.0 & & & & \\
$\mathit{1957.496}$ & $\mathit{346.7}$ & & $\mathit{0.91}$ & & $\mathit{4}$
	& $\mathit{6.0}$ & $\mathit{8.1}$ & & & & {\em m} \\
\hline
1960.78   & 346.2 & & 0.91 & & 5 & & & & & & H61 \\
\hline
& & & 0.9 & & 1 & & & 1.98 & & & vH66 \\
\hline
1991.25 &349&& $0.929\pm0.003$ & & & & & 1.86 & & & E97 \\
& $346$ && $0.9$ & & & 6.1 & 8.1 & & & & DN00 \\
& $349$ & & & $16.8$ & & 6.1 & & & 14.1 & & DN00 \\
& $352.5$ && $0.95$ & & & 6.19\tablefootmark{H} & 8.02\tablefootmark{H} & & & & FM00 \\
\hline
1996.7013 & 349.5 & & 0.915 & & 1 & & & & & & H00 \\
\hline
1995.47 & 350.5 & & 1. & & 1 & & & 2.2 & & & A98 \\
1996.74 & 352 & & 0.96 & & 3 & & & 2.1 & & & A98 \\
\hline
1998.789 & 349.0 & & 0.90 & & & 6.0 & 7.9 & & & & D00 \\
\hline
2000.44 & & 340.1 & & 17.16 & & & & & & & W06 \\
\hline
2000.5000 & 349.7 & & 0.907 & & 1 & & & & & & D04 \\
2000.5002 & 349.8 & & 0.911 & & 1 & & & & & & \\
\hline
2001.365 & & 339.8$\pm$0.1 & & 16.572$\pm$0.113 & 2 & 7.700$\pm$0.340 & & &
	13.950$\pm$0.06 & & H13 \\
\hline
2003.788  & $348.9$ & & $0.89$ & & 6 & & & & & & M04 \\
\hline
2004.878  & $349.4\pm0.007$ & & $0.919\pm0.007$ & & 1 & & & & & & S06 \\
\hline
\end{tabular}
\end{center}
\tablefoot{$n$: number of measurements;
\tablefoottext{D}{observations by Dembowski};
\tablefoottext{H}{Hipparcos $V_T$ magnitude}.
Additional measurement of positions was reported for 2009 in the current
version of the Washington Double Star Catalog, but no further details
were found.
}

\tablebib{
(B73) \cite{Burnham:1873}: micrometer, Chicago%?
, 6-inch Alvan Clark telescope;
(B74) \citep{Burnham:1874}: micrometer, Washington, 26-inch telescope;
(P51) \cite{Pretre:1951}: micrometer, Pic du Midi, 38-cm;
(vdB58) \cite{vandenBos:1958}: micrometer, Lick, 12-inch reflector;
(H61) \cite{Heintz:1961}: micrometer, M\"unchen, 28.5cm refractor;
(vH66) \cite{vanHerk:1966}: micrometer, Observatoire de Nice, 38-cm;
(E97) \cite{ESA:1997}: HIPPARCOS;
(DN00) \cite{Dommanget:Nys:2000}: HIPPARCOS;
(FM00) \cite{Fabricius:Makarov:2000}: HIPPARCOS;
(H00) \cite{Hartkopf:etal:2000}: speckle, Mount Wilson, 2.5-m telescope;
(A98) \cite{Alzner:1998}: micrometer, Hemhofen,
32.5-cm Cassegrain and 360-mm Zeiss-Newtonian;
(D00) \cite{Douglass:etal:2000}: micrometer, Washington, 66-cm refractor;
(W06) \cite{Wycoff:etal:2006}: from 2MASS observations using data mining;
(D04) \cite{Docobo:etal:2004}: speckle, Calar Alto, 1.52-m telescope;
(H13) \cite{Hartkopf:etal:2013}: 20-cm, UCAC R$_u$;
(M04) \cite{Mason:etal:10:2004}: speckle, USNO Naval, 26-inch refractor;
(S06) \cite{Scardia:etal:2006}: speckle, Brera, Merate, 1-m Zeiss telescope;
{\em m}: mean values ({\em in italics})
}
\end{sidewaystable*}

\end{document}